\documentclass[aj_pt4]{aastex}

\bibpunct{(}{)}{;}{a}{}{;}

\begin{document}
\slugcomment{Accepted for publication in AJ; to appear April 2001}

\shortauthors{Adams et al. }
\shorttitle{Mass and Structure of the Pleiades}

\title{The Mass and Structure of the Pleiades Star Cluster from 2MASS}

\author{Joseph D. Adams\altaffilmark{1,2}}
\affil{University of Massachusetts, Department of Astronomy, LGRT 531-A, Amherst, MA 01003}
\author{John R. Stauffer\altaffilmark{3}}
\affil{Harvard-Smithsonian Center for Astrophysics, 60 Garden Street, Cambridge, MA 01238}
\author{David G. Monet}
\affil{U.S. Naval Observatory Flagstaff Station, P.O. Box 1149, Flagstaff, AZ 86002-1149}
\email{dgm@nofs.navy.mil}
\author{Michael F. Skrutskie}
\affil{University of Massachusetts, Department of Astronomy, LGRT 532, Amherst, MA 01003}
\email{skrutski@astro.umass.edu}
\and
\author{Charles A. Beichman}
\affil{Jet Propulsion Laboratory, California Institute of Technology, JPL 180-703, Pasadena, CA 91109}
\email{chas@pop.jpl.nasa.gov}

\altaffiltext{1}{Visiting Astronomer, Kitt Peak National Observatory, National Optical Astronomy Observatories, which is operated by the Association of Universities for Research in Astronomy, Inc. (AURA) under cooperative agreement with the National Science Foundation.}
\altaffiltext{2}{Present address: Boston University, Institute for Astrophysical Research, 725 Commonwealth Avenue, Boston, MA 02215; e-mail: jdadams@bu.edu}
\altaffiltext{3}{Present address: Infrared Processing and Analysis Center, California Institute of Technology, Mail Code 100-22, Pasadena, CA 91125; e-mail: stauffer@ipac.caltech.edu}

\begin{abstract}
We present the results of a large scale search for new members of 
the Pleiades star cluster using 2MASS near-infrared photometry and 
proper motions derived from POSS plates digitized by the USNO PMM
program. The search extends to a 
$10^\circ$ radius around the cluster, well beyond the presumed tidal 
radius, to a limiting magnitude of $R \sim 20$, corresponding to
$\sim$ 0.07 M$_\odot$ at the distance and age of the Pleiades.
Multi-object spectroscopy for 528 candidates
verifies that the search was extremely effective at detecting cluster stars 
in the 1 - 0.1 M$_\odot$ mass range using the distribution of H$\alpha$ 
emission strengths as an estimate of sample contamination by field stars. 

When combined with previously identified, higher mass stars, this search
provides a sensitive measurement of the stellar mass function and
dynamical structure of the Pleiades. The degree of tidal elongation of 
the halo agrees well with current $N$ body simulation results.
Tidal truncation affects masses below $\sim 1$ M$_\odot$.
The cluster contains a total mass $\sim 800$ M$_\odot$.
Evidence for a flatter mass function in the core than in 
the halo indicates the depletion of stars in the core with mass less than
$\sim 0.5$ M$_\odot$, relative to stars with mass $\sim 1 - 0.5$ M$_\odot$,
and implies a preference for very low mass objects to populate the halo or 
escape. The overall mass function is best fitted with a lognormal
form that becomes flat at $\sim$ 0.1 M$_\odot$. Whether sufficient
dynamical evaporation has occurred to detectably flatten the initial
mass function, via preferential escape of very low mass stars and brown 
dwarfs, is undetermined, pending better membership information for stars
at large radial distances. 

\end{abstract}


\section{INTRODUCTION}

The age distribution of open clusters suggests they are transient objects 
which are dissolved by the Galactic tidal field in time scales typically 
less than $\sim 1$ Gyr \citep{wielen71}. Due to their small number of constituent 
stars, open clusters have short relaxation times of order
$10^8$ years or less, comparable to their typical crossing time, and can thus be 
understood only as collisional systems which evolve rapidly compared to 
larger systems, such as globular clusters \citep{king80}. This rapid 
evolution produces observable effects, such as mass segregation and the 
development of a halo, where tidal 
forces cause the dispersal of escaping stars. Despite the theoretical
problems associated with collisional systems, the small number
density of stars in open clusters allows the use of direct $N$ body 
simulation as a practical tool of study \citep[eg.][]{terlevich87, 
kroupa95, aarseth99, portegies00}. Unfortunately, $N$ body simulations have
suffered from a lack of detailed observational guidance.

The Pleiades cluster represents an appropriate laboratory for stellar 
dynamics -- bright and rich in stars, but old enough to be dynamically 
evolved. Studies of the Pleiades date back to \citet{trumpler20} 
and \citet{hertzsprung47} (HII), and include those of \citet{jones81}, 
\citet{haro82} (HCG), \citet{vanLeeuwen83}, \citet{stauffer91} (SK), 
\citet{rosvick92}, and
\citet{schilbach95}. The last large, faint proper 
motion survey was that of \citet{HHJ93} (HHJ) 
which revealed 440 probable members in 25 square degrees around 
the cluster center. Although the HHJ survey discovered many low 
mass cluster members, it was spatially incomplete, and the 
authors suggested that a larger survey would be fruitful. 
Research groups during the last decade have conducted many deep photometric 
searches for brown 
dwarfs in small fields aimed at studying the substellar mass function,
such as \citet{stauffer89, stauffer94, ZO97, bouvier98}, and  \citet{festin98}. 
Most have concluded that brown dwarfs are abundant in the Pleiades,
but not numerous enough to contribute significantly to the total
cluster mass. Observations of lithium absorption in brown 
dwarfs has lead to a precise age measurement for the Pleiades
of 120 Myr \citep{stauffer98}, a result which has interesting
implications for the dynamical state of the cluster 
as well as stellar astrophysics.

Despite a rich history of vigorous research in the Pleiades,
the spatial incompleteness of previous surveys leaves some basic
but critical properties of the cluster poorly constrained, such as its 
overall extent, structure, mass function and total mass. This paper 
summarizes the results of a spatially complete photometric and proper 
motion search in the Pleiades using the Two Micron All Sky Survey
(2MASS) and proper motions determined from First Palomar
Observatory Sky Survey (POSS I) plates and the Second Palomar Observatory
Sky Survey (POSS II) \citep{reid91} plates. Our primary goal is to
characterize the fundamental structure of the Pleiades and
estimate the mass function below 0.5 M$_\odot$ to the completeness of 
the POSS $E$ plates, roughly 0.1 M$_\odot$ for cluster members.

\section{DATA}

This study used 2MASS point source data in the Pleiades field, 
which provided $JHK_s$ photometry and positions on a uniform 
reference frame, and also indicated the presence of any optical counterpart in 
the USNO-A catalog \citep{monet96}. The 2MASS data exceeded a signal-to-noise 
ratio of 10 at $J=15.8$, $H=15.1$, and $K_s=14.3$. The Pleiades area 
covered a $10^\circ$ radius around the nominal center of 
$\rm{RA}=3^{\rm{h}}47^{\rm{m}}$, $\rm{DEC}=24^\circ7^\prime$ 
\citep{lynga87}. 2MASS detected $\sim 7.8 \times 10^5$ sources 
in $JHK_s$ in the field down to $K_s = 15$ (compare with several hundred known
Pleiades members). 

The USNO's Precision Measuring Machine (PMM) program \citep{monet98}
scanned and digitized POSS I $E$ plates and POSS II $F$ plates.
The PMM detections provided first and second epoch point source positions 
and instrumental $R_E$ and $R_F$ magnitudes. Errors in positions and magnitudes
should be comparable to the USNO-A catalog, about $0.25^{\prime\prime}$ and
0.25 magnitudes respectively, in the magnitude range 12 -- 19 \citep{monet96}.
The $\sim 35$ year baseline between POSS I and POSS II allowed proper motion 
measurements to distinguish faint, red Pleiades candidates from field stars.
We correlated  $E$ and $F$ detections in the Pleiades field using a 
$5^{\prime\prime}$ search radius over roughly a $10^\circ$ radius around 
the nominal cluster center. Before computing proper motions, 
we attempted a first-order correction for systematic dispersion 
of the POSS I - POSS II positional offsets over large plate 
areas and across plate boundaries by subtracting the mean positional
offsets between POSS I and POSS II counterparts
brighter than $R_E=18$ in $10^\prime \times 10^\prime$ boxes.
A correlation of POSS sources with 2MASS $JHK_s$
detections produced a database of $\sim 6.3 \times 10^5$ sources with 
2MASS positions, $R_ER_FJHK$ magnitudes, 
and relative proper motions. The correlation with 2MASS 
and the five band detection requirement ensures that the database
contains few spurious sources.

\section{CLUSTER EXTRACTION}

\subsection{Color selection}

\begin{figure}
\plotone{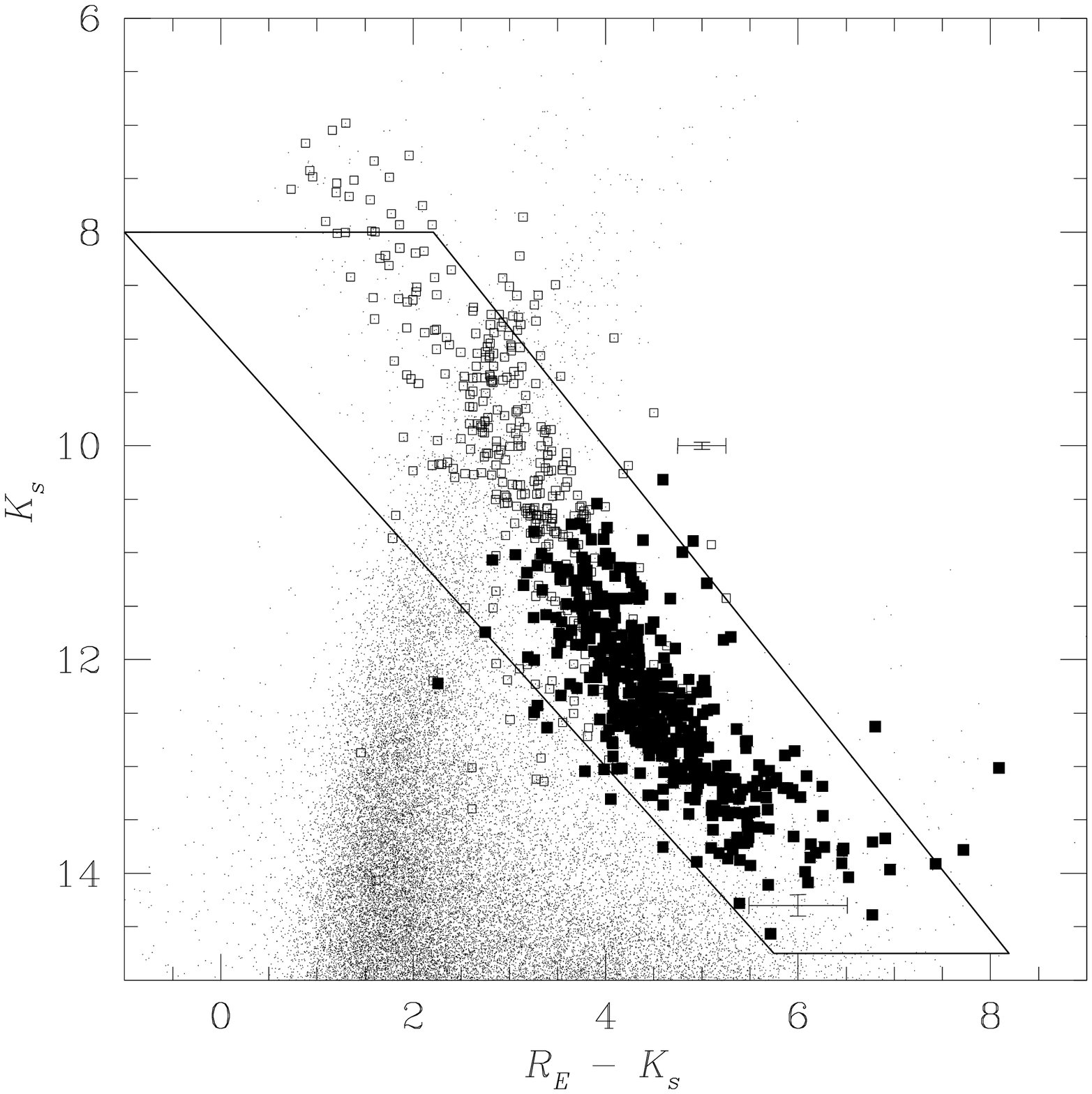}
\caption{Color-magnitude diagram using instrumental
$R_E$ and 2MASS $K_s$ magnitudes for field stars in a small
area ($\sim$ 12 deg$^2$) around the Pleiades. The open squares
represent colors for a sample of previously published Pleiades 
candidates brighter than the HHJ sample,
while the solid squares represent colors for HHJ stars. Typical error
bars are shown along the main sequence near $K_s = 10$ and $K_s = 14$.
The boxed area indicates the region in which sources were selected for 
proper motion analysis.
\label{fig:colorkrk}}
\end{figure}

\begin{figure}
\plotone{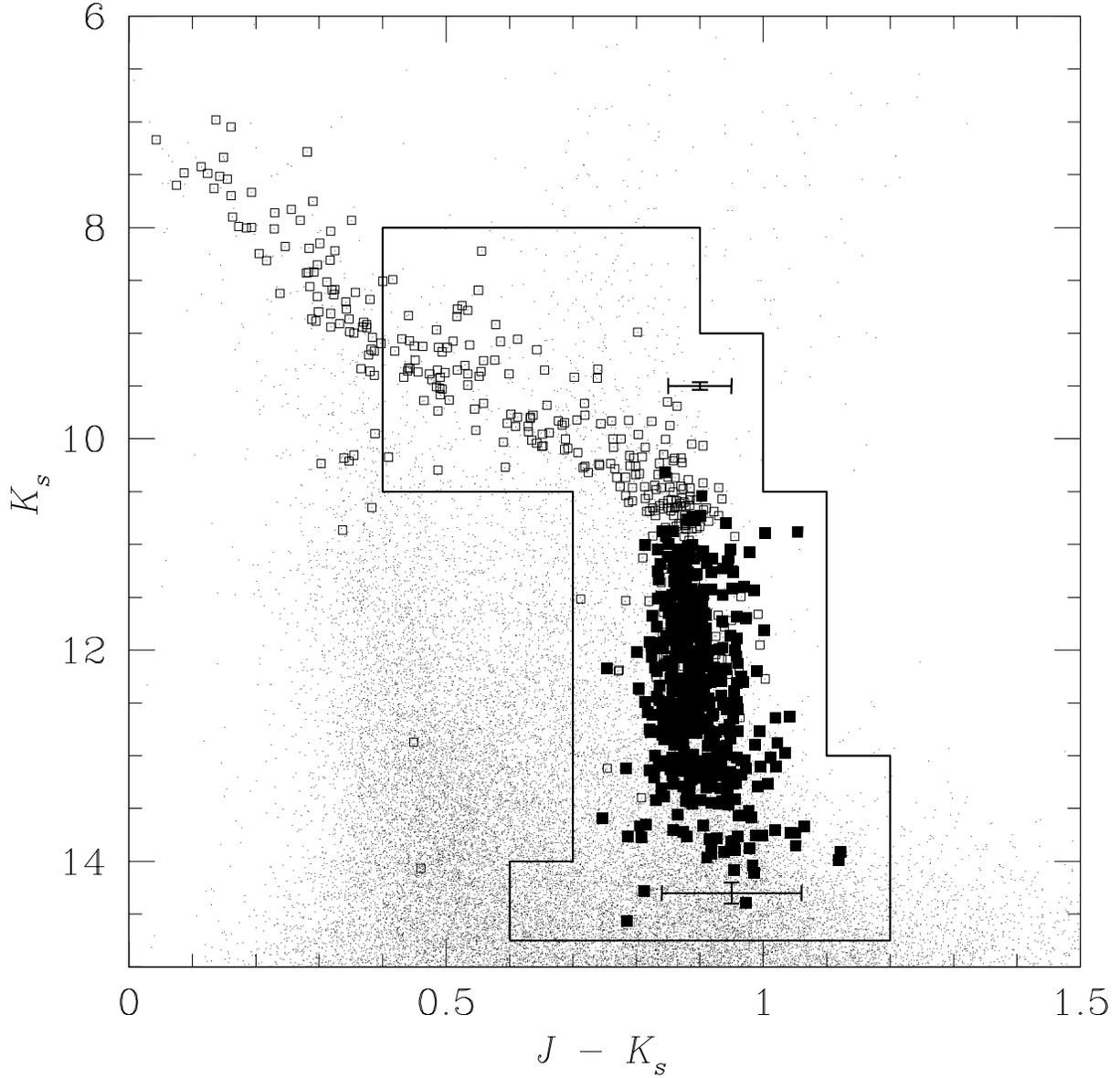}
\caption{2MASS $K_s:J-K_s$ color-magnitude diagram
for the sources described in Figure \ref{fig:colorkrk}. Symbols
and boxed region analogous to those in Figure \ref{fig:colorkrk}. 
\label{fig:colorkjk}}
\end{figure}

\begin{figure}
\plotone{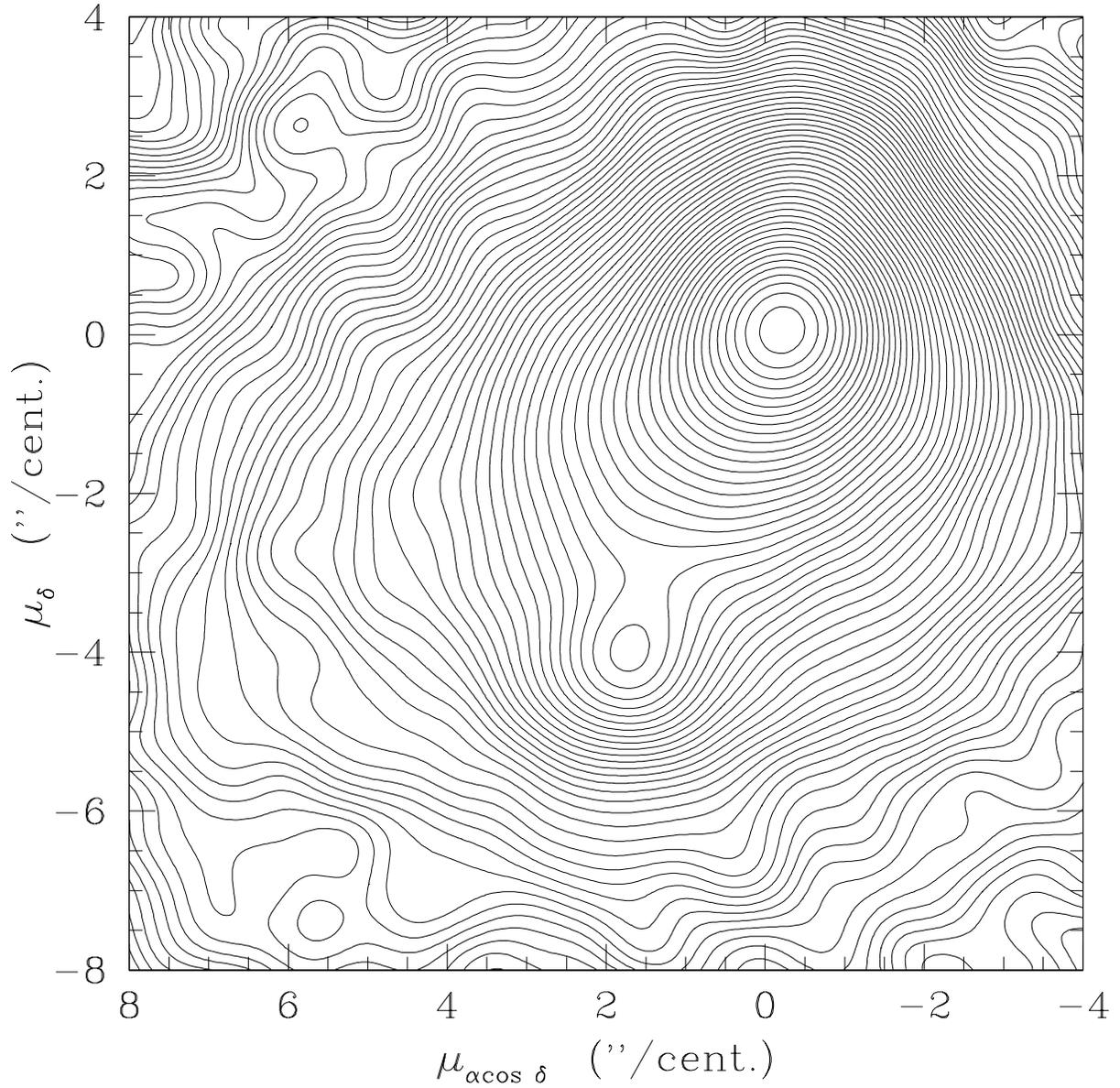}
\caption{Contours derived from a Hess diagram of
proper motion vectors, after color selection, for
sources brighter than $K_s=14$. The center of the field star 
distribution lies near (0,0) $^{\prime\prime}$/cent. and the 
cluster distribution is visible near (2,-4) $^{\prime\prime}$/cent. 
\label{fig:pmhess}}
\end{figure}

Even in proper motion space the large area of this Pleiades search
introduces a substantial amount of field star contamination. 
To reduce this contamination, a broad color filter selected only
sources with colors similar to known Pleiades members. We chose to 
include sources in 
the range $8.0 \le K_s \le 14.75$ based on the dynamic range of 
the PMM extractions at the bright end and by the sensitivity of 
the POSS plates at the faint end. The color filter
included the faint portion of the Pleiades locus, 
in the range $10 \le K_s \le 14$, composed mainly of late K and
early to mid-M dwarfs. We extended the faint end of the 
magnitude range beyond $K_s = 14$ to try to include objects near 
the hydrogen burning limit at $K_s \approx 14.5$.
Figures \ref{fig:colorkrk} and \ref{fig:colorkjk} show the specific 
location of color selection in the $K_s$ {\it vs.} $R_E-K_s$ 
and $J - K_s$ color-magnitude 
diagrams, respectively. We deliberately chose
broad regions to account for dispersion due to photometric error
at faint magnitudes, as well as binarity. 
For sources fainter than $K_s = 14$, we excluded 2MASS sources that have
a USNO-A counterpart in the 2MASS database, since faint, red 
Pleiades members will likely be absent in the blue-limited USNO-A 
catalog. In total, $\sim 38000$ sources to $K_s=14$ and $\sim 5700$ sources 
fainter than $K_s=14$ survived the
color selection. In Figure \ref{fig:pmhess}, the density of points
in a vector point diagram (VPD) for these color-selected stars
reveals the cluster in proper motion space, as well as the still-dominant
field star population. The proper motion dispersion due to internal
dispersion in velocities, $\sim 0.5$ km/s \citep{jones70}, is negligible
compared with the uncertainty in the astrometry. 

\subsection{Membership probability}

The color cuts sufficiently reduce the density of 
field-to-cluster sources in proper motion space to allow
the computation of relative membership probabilities 
$p$ given by
\begin{displaymath}
p = \frac{\Phi_c}{\Phi_c + \Phi_f}
\end{displaymath}
where $\Phi_c$ and $\Phi_f$ are the distribution functions
for the cluster and field stars respectively in the VPD.
We fitted the distribution functions using a modified version of 
the maximum likelihood technique of \citet{sanders71}.
The distribution functions resembled those discussed in detail 
by \citet{jones91}, ie., a bivariate
gaussian for the cluster, and an exponential distribution along the
cluster's mean proper motion with a gaussian in the orthogonal
direction for the field. However, we chose not to
include radial distance from the center of the cluster 
as a variable parameter in the distribution function in 
order to measure
the true cluster extent free of model bias and to estimate the
surface density of non-member stars with coincidentally 
high membership probabilities at large radial distance.

The distribution function fitting procedure operated on 
$\sim$ 5600 sources contained in a region approximately 
$5 \times 5$ $^{\prime\prime}$/cent. around the cluster centroid.
Due to the large number of field stars in the region, we found 
fitting the distribution functions in small magnitude 
groups difficult, particularly for faint stars. We fitted the distribution
function without a magnitude dependence.
This will affect the membership probabilities by favoring 
parameters that best fit the magnitudes with the least 
field contamination. However, the effect should be small except 
at the faint end of the sample. Membership probabilities
for the $\sim 500$ sources fainter than about $K_s = 14$ are much less
reliable than those for the the relatively brighter stars for two reasons. 
First, degradation in the PMM astrometry at faint magnitudes 
causes a larger dispersion in proper motion space, which makes
modelling their distribution difficult. The faint clusters members
are dispersed to smaller membership probabilities 
than relatively brighter cluster members. Second, field 
contamination increases at faint magnitudes. Membership of candidates 
fainter than about $K_s \approx 14$ will thus rely more heavily on 
their spectral properties.

\begin{figure}
\plotone{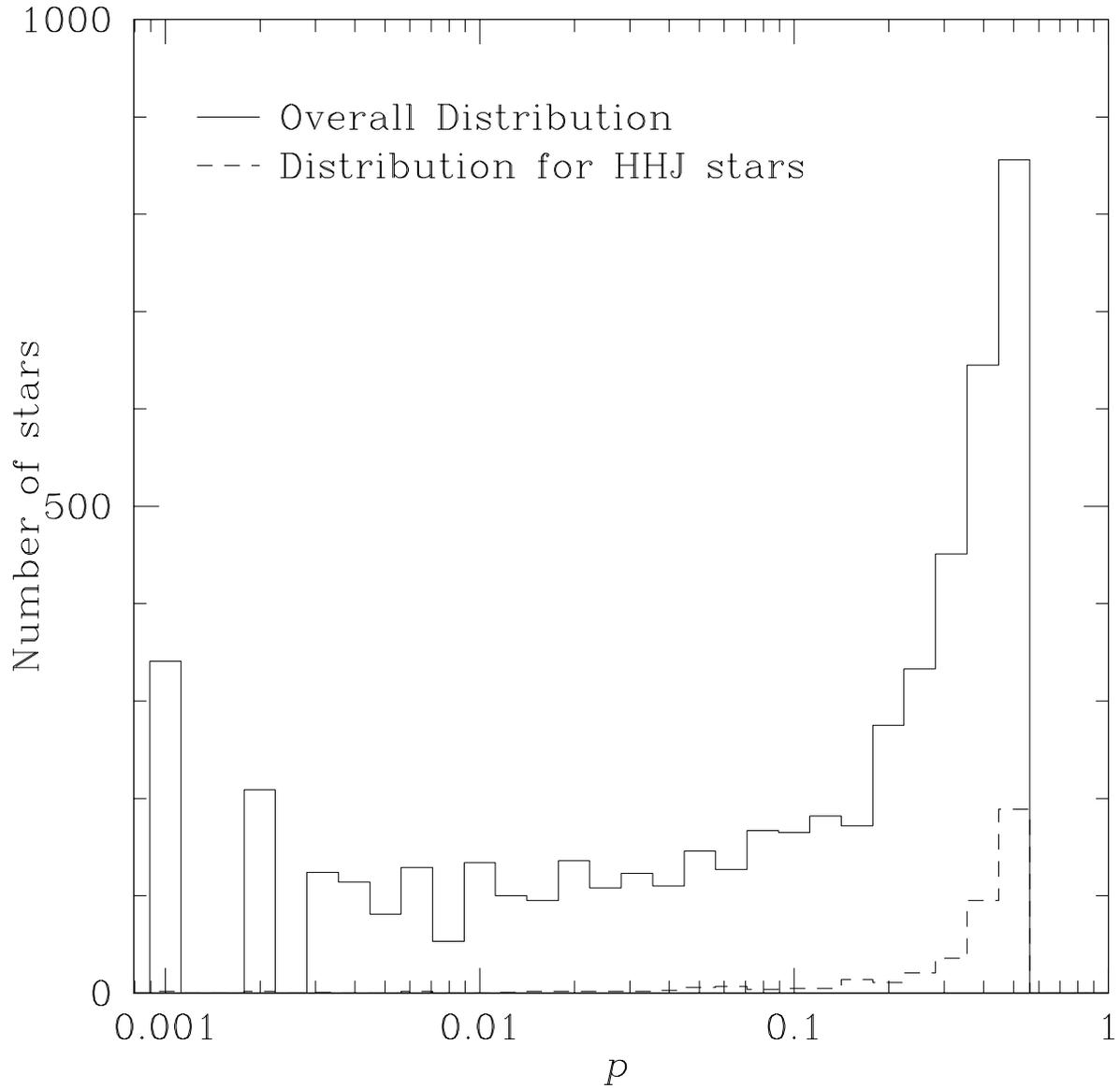}
\caption{Distribution of membership probabilities $p$ for Pleiades 
candidates. The dotted line shows the $p$ distribution for
recovered HHJ stars.
\label{fig:phist}}
\end{figure}

\begin{figure}
\plotone{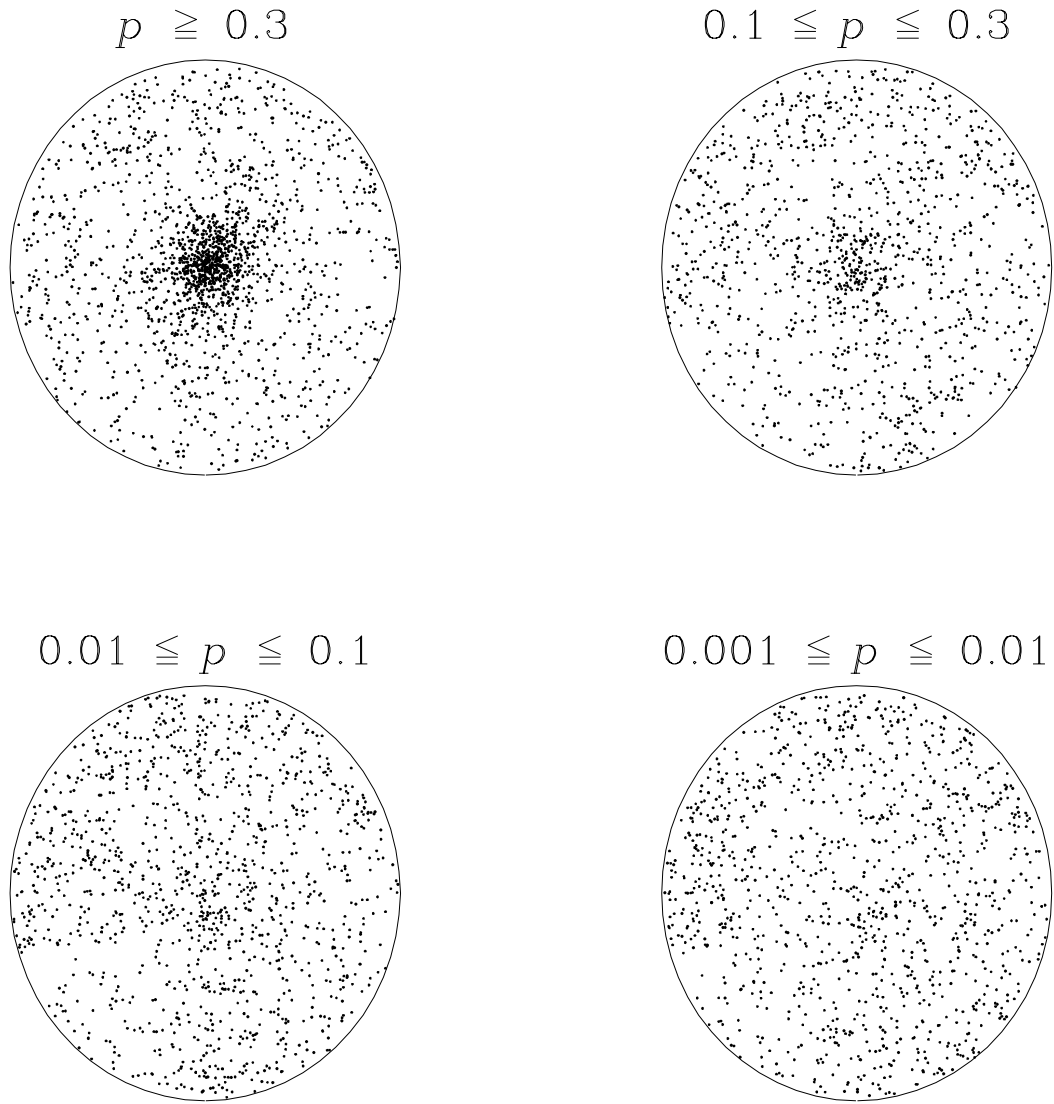}
\caption{Spatial distribution of Pleiades candidates in the probability 
ranges $p > 0.3$; $0.1 \le p < 0.3$; $0.01 \le p < 0.1$; 
and $0.001 \le p < 0.01$. The field apertures are each $10^\circ$ in
radius. Left corresponds to the eastern direction.
\label{fig:plpos4}}
\end{figure}

The large area covered by this work results in
enough field contamination to numerically dilute the membership
probabilities. However, Figure \ref{fig:phist} shows that most 
sources with $p > 0.3$ are likely to be cluster members based 
on the overall distribution of probabilities. The spatial distribution 
of sources with membership probability depicted in Figure 
\ref{fig:plpos4} demonstrates qualitative statistical
consistency of proper motion membership probability with true 
association with the cluster.

\section{SPECTROSCOPY}

In order to determine the success of the proper motion search, 
and estimate the field star contamination in our sample, we obtained 
optical spectra for a large number of our Pleiades 
candidates. M dwarf members of the Pleiades exhibit strong 
chromospheric activity \citep{prosser91, steele95}, detectable through
resulting H$\alpha$ emission, while field M dwarfs are usually less 
active \citep{hawley96}. Therefore, absence of H$\alpha$ emission 
distinguishes obvious field stars from Pleiades members.

The spectra were taken using the Hydra red cable fiber-fed
spectrograph and T2K CCD camera at the WIYN telescope\footnote{The WIYN Observatory is a joint facility of the University of Wisconsin-Madison, Indiana University, Yale University, and the National Optical Astronomy Observatories.}
\citep{barden94} at Kitt Peak National Observatory during 
November 25-28, 1999. 
The spectrograph was configured with the 600@13.9 grating in 
the 6200 - 8900 {\AA} wavelength region for a dispersion of about 
1.4 \AA/pixel. While most 
spectroscopic fields fell within $2^\circ$ of the cluster 
center, we targeted several fields at larger radial distance.
Table \ref{tab:wiynfields} lists the positions of the centers of
the WIYN Hydra fields, each approximately
1$^\circ$ in diameter, along with the number of candidates and 
control stars observed, and total exposure time used in each field.
Fields near the center of the cluster partially overlapped so 
that about 128 candidates were observed more than once. The observations
included 121 control targets with colors similar to the
Pleiades candidates but that were expected to be field stars 
based on having proper motion farther than 3 $^{\prime\prime}$/cent. 
from the cluster centroid in the VPD.

\begin{figure}
\plotone{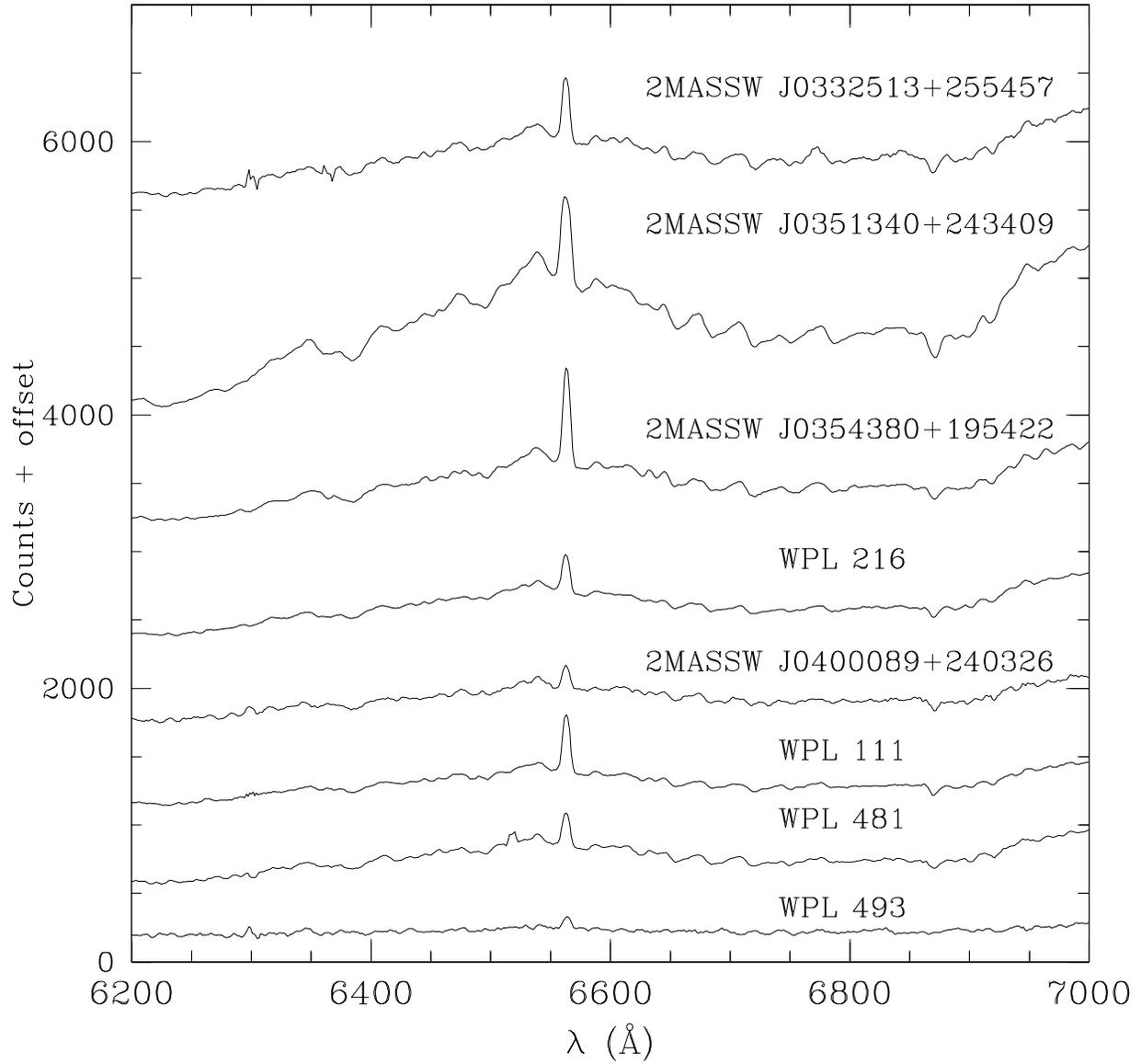}
\caption{Smoothed spectra for representative Pleiades candidates that 
show H$\alpha$ emission. The WPL identifications are adopted from 
\citet{adams00}.
\label{fig:plspec}}
\end{figure}

We used standard CCD reductions and analysis routines in 
IRAF\footnote{IRAF is distributed by the 
National Optical Astronomy Observatories, which is operated by the
Association of Universities for Research in Astronomy, Inc., under
contract with the National Science Foundation.}, 
and susequently measured the equivalent width of H$\alpha$ line 
($W_{H\alpha}$) from each one-dimensional spectrum using the
FITPROFS routine over a spectral region of $6555 - 6569$ {\AA}. For 
candidates observed more than once, we averaged the individual 
$W_{H\alpha}$ measurements. Figure \ref{fig:plspec} displays
a small sample of spectra for candidates with H$\alpha$ emission.

\begin{figure}
\plotone{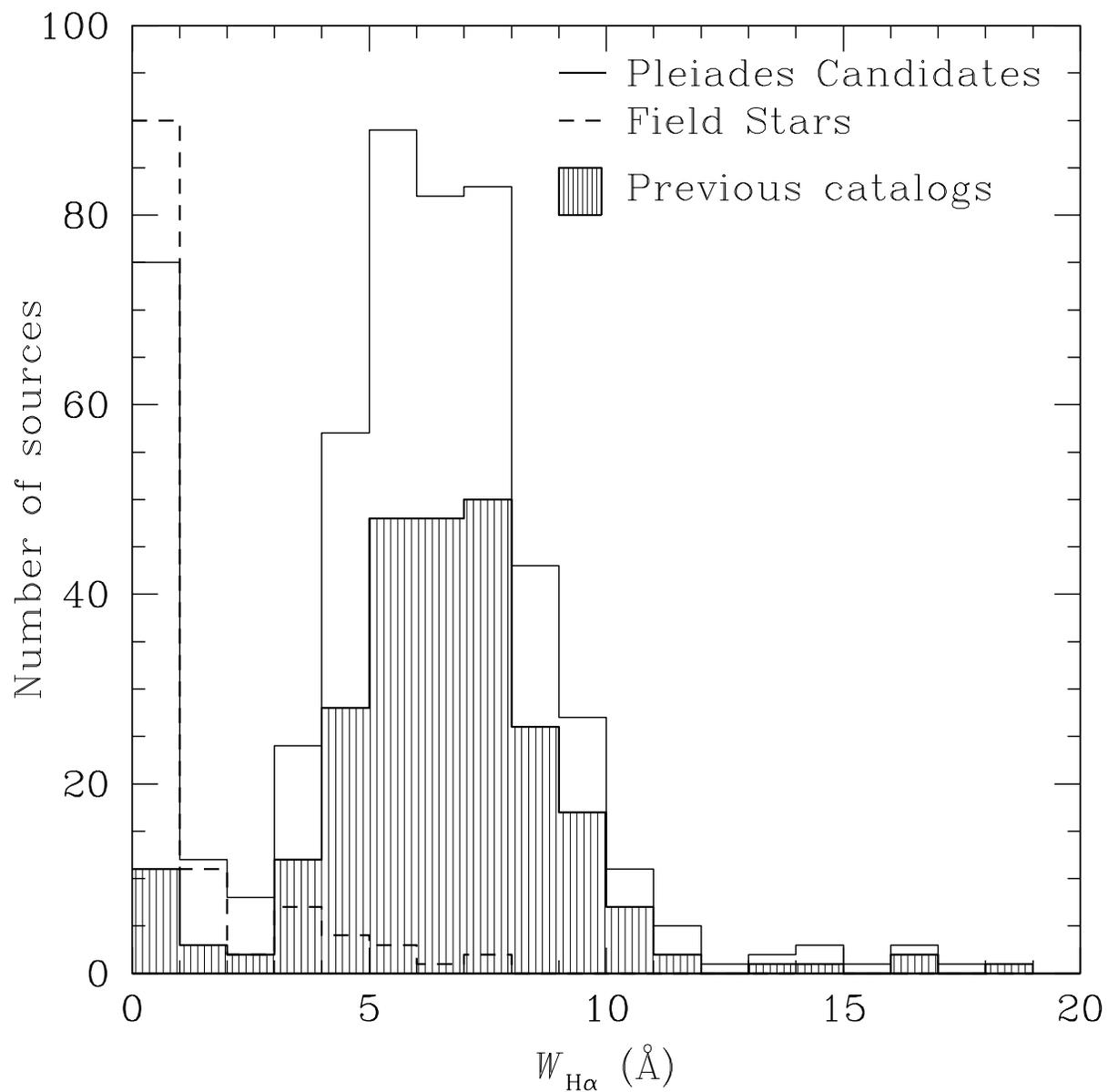}
\caption{Histogram of H$\alpha$ equivalent widths $W_{H\alpha}$ for all sources 
in the Pleiades spectroscopic study. The shaded region indicates the 
fraction of candidates identified by previous Pleiades studies. The 
dashed line represents the distribution for the control sample.
\label{fig:Hhist}}
\end{figure}

Figure \ref{fig:Hhist} shows the distribution of $W_{H\alpha}$ 
for the 528
Pleiades candidates as well as the 121 control sources
in the spectroscopic study. The shaded region indicates our 
$W_{H\alpha}$ measurement for previously known members in the 
sample. A $\chi^2$ test suggests there is less than $10^{-5}$ 
probability that the candidate sample and the control sample are 
derived from the same distribution function. However, about 
75 Pleiades candidates in the spectroscopic sample do not show 
strong H$\alpha$ emission, indicating that a fraction of our overall 
sample contains field stars.

\section{MEMBERSHIP LIST}

In total, this Pleiades search has identified 4233 
possible Pleiades members with $p \ge 0.01$. Most will actually 
be non-members at large radial distances.
Roughly 1200 are high probability members with $p \ge 0.3$ within
a $6^\circ$ distance from the cluster center. 

\subsection{Contamination \label{sec:contam}}

Table \ref{tab:Hstats} shows the statistical dependence of
$W_{H\alpha}$ with respect to proper motion membership probability, 
$K_s$ magnitude, and radial distance. As expected, the contamination
is worst at large radial distance and at faint magnitudes. 
For sources with $p \ge 0.01$ in the magnitude range 
$11 \le K_s \le 14$ within
$5^\circ$ of the cluster center, we estimate the overall
contamination to be about 13\%. The lower surface 
density of field stars at relatively brighter magnitudes 
implies a lower contamination level for brighter 
stars. Outside a $5^\circ$ radius, we expect the contamination to dominate the
sample, but cannot yet quantify it without individual membership
information obtained, for example, from radial velocity measurements.

The sample of 121 field stars contained 104 M dwarfs that had spectral types 
similar to the Pleiades candidates, M2 -- M6 \citep{adams00}. Among these 
104 objects, 29 had $W_{H\alpha} > 1$ \AA, and 18 had $W_{H\alpha} > 3$ \AA. 
Assuming that stars with $W_{H\alpha} > 1$ {\AA}
are indeed dMe stars, the 28\% dMe rate in the M2 -- M6 range
agrees well with the data given in \citet{hawley96},
although the Pleiades field sample is too 
small to determine the dMe fraction as a function of spectral type.
Thus, from the 70 Pleiades candidates
in the range M2 -- M6 that we assume are not members ($W_{H\alpha} < 1$ {\AA}), 
we can estimate roughly 27 dMe stars in the sample of 434 with 
$W_{H\alpha} > 1$ {\AA}, or 6\% dMe contamination.

\subsection{Completeness and sensitivity}

The dynamic range of the PMM extractions limits the completeness 
at the bright end of this search. Counts for sources with correlated $R_ER_FJHK_s$
detections indicate that the PMM detections are complete up to 
$R \approx 8$, corresponding to $\sim$ 1.5 M$_\odot$ for Pleiades
members. Since our primary goal is to detect low mass stars, incompleteness
at the bright end of our sample should not pose a significant problem
due to the large amount of published data for bright Pleiades members.

This work recovered over 92\% of the HHJ sample including
several of their faintest objects, such as HHJ 3 and HHJ 2 
(albeit at relatively low membership probability), indicating that 
it is relatively complete to $R=19$. This work has also discovered
about 269 candidates with $p \ge 0.3$ that were in the HHJ 
survey area but not in their list. Spectra for 109 of these candidates
show that 83\% have $W_{H\alpha}$ consistent with cluster membership 
(3 -- 14 \AA), while 94\% of HHJ stars in our list have $W_{H\alpha}$ 
consistent with cluster membership. Thus, while our study appears to 
be more complete than HHJ, it is also more contaminated. If 83\% of the
269 new objects in the HHJ and region and sensitivity range are in fact 
Pleiades members, then the HHJ survey is 66\% complete relative to a projected
total of 663 objects, in good agreement with their completeness estimation
of 70\% through most of their sensitivity range.

This search also recovered 8 objects from the survey of \citet{bouvier98},
including CFHT-PL-5 and CFHT-PL-10, 3 PPl objects from 
\citet{stauffer89} including PPl 13, and no objects from \citet{ZO97}.
Thus it is very incomplete at the hydrogen burning 
limit, $R = 20$ \citep{stauffer98}, and below. 

\subsection{Merging Pleiades catalogs \label{sec:merge}}

In order to construct a ``complete'' Pleaides catalog containing faint stars
from this work and brighter stars from previous work, we merged our
list of Pleiades candidates with probable members contained in the CfA Open 
Cluster Database\footnote{Provided by C.F. Prosser (deceased) and J.R. Stauffer, 
and which currently may be accessed at http://cfa-ftp.harvard.edu/~stauffer/, 
or by anonymous ftp to cfa-ftp.harvard.edu, cd /pub/stauffer/clusters/}, 
derived primarily from HII, SK, HCG, and HHJ. 
We did not merge our list with any other external catalogs containing 
members brighter than our sample, since most are spatially incomplete.

For this study, approximate masses are suitable for dividing all Pleiades 
candidates into dynamical mass groups. Adjustment of apparent magnitudes,
using a distance modulus of 5.5 \citep{lynga87}, yielded absolute 
magnitudes. Interpolation of mass-luminosity relations from \citet{bcah98} 
provided masses for the candidates detected in this study, using 2MASS $J$ 
magnitudes or HHJ $I$ magnitudes for a few HHJ stars not recovered. 
Mass-$M_V$ relations from \citet{henry93} for 
$M_V \ge 1.45$, and \citet{allen73} for $M_V < 1.45$, converted $V$
magnitudes into masses for stars in the CfA Open Cluster Database.
Variations between photometric systems of the mass-luminosity relations
and the data will not affect the stellar mass estimates appreciably.

\section{RESULTS AND DISCUSSION}

\subsection{Mass segregation and the mass function}\label{sec:mseg}

\subsubsection{Internal Structure}

The long term internal evolution of an open cluster is described 
in terms of energy equipartition from kinetic theory of collisional systems. 
Briefly, lower mass stars, through encounters with high mass stars and
binaries systems, acquire higher velocities resulting in larger, more 
radial orbits. High mass stars lose velocity and thus rapidly ``sink'' 
to the center of the cluster, where they can form binaries which 
become tighter, or disrupt, as a result of the encounters in the core. 
Mass segregation has been discussed by \citet{spitzer75},
and verified by the $N$ body simulations of \citet{terlevich87} and 
\citet{kroupa95}. This section discusses radial distribution of mass in the 
Pleiades, emphasizing stars with mass less than 1 M$_\odot$.

\begin{figure}
\plotone{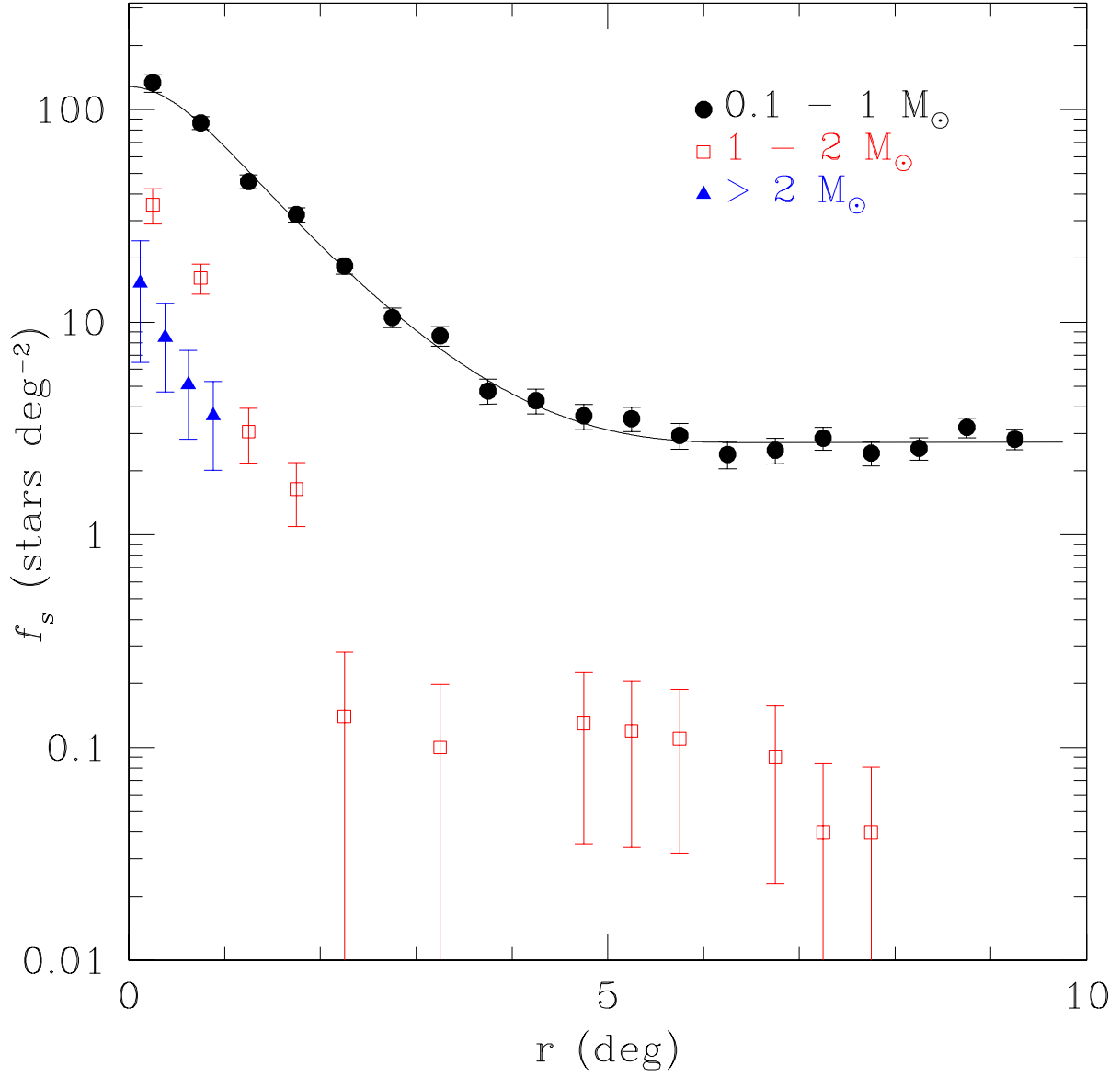}
\caption{Radial profiles for surface density $f_s$, binned into annuli around 
the nominal center of the Pleiades, for three mass groups. 
The solid line shows the best 
fit single-mass King profile for the mass group 0.1 -- 1 M$_\odot$, 
superimposed on the mean background surface density. Error bars represent 
Poisson uncertainties in each bin.
\label{fig:fsall}}
\end{figure}

To estimate the extent of the cluster on the sky, we divided Pleiades 
candidates with $p \ge 0.3$ into high ($\ge 2$ M$_\odot$), 
intermediate (1 -- 2 M$_\odot$), and low ($< 1$ M$_\odot$) mass 
groups and then binned the high mass stars into 
$0.25^\circ$ annuli, and the low and intermediate mass stars into
$0.5^\circ$ annuli. Figure \ref{fig:fsall} shows the radial distribution 
of surface number density for the three mass groups
out to a radius $r \approx 10^\circ$ from the cluster center.
The low mass group clearly fits a standard, three-parameter,
single-mass King model \citep{king62}.
Subtraction of the mean background surface number density,
determined in the annulus at 7 -- 10$^\circ$, from each annulus yielded an
estimate of the cluster surface number density.
The best chi-square fit gave a core radius $r_c = 1.0^\circ - 1.3^\circ$ and 
King tidal radius $r_t = 5.8^\circ - 6.8^\circ$ at 99\% confidence.
Figure \ref{fig:fsall} shows this fit superimposed on the mean background 
surface density. Note that these numbers only parameterize the empirical 
distribution of low mass stars on the sky, and this King tidal 
radius does not necessarily correspond to the location of the cluster's 
Lagrangian points along the Galactic $X$ direction. Since the cluster's location  
is $l=166^\circ$ and $b=-23^\circ$, the Lagrangian points in the 
$X$ direction lie nearly along the line of sight, projected near 
the dense cluster center.

\begin{figure}
\plotone{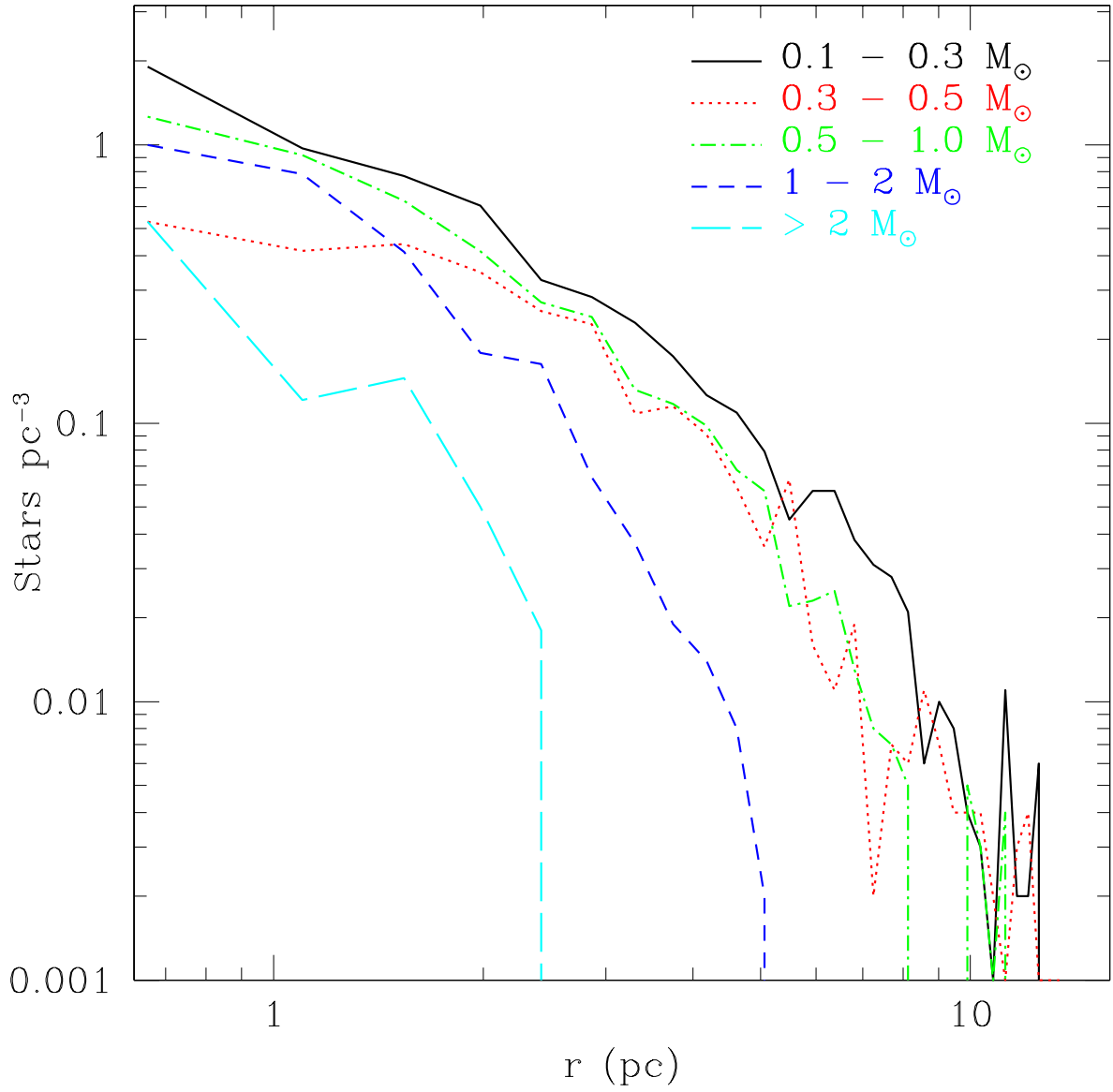}
\caption{Approximate spatial density for Pleiades stars of
different mass groups derived from a King model. 
\label{fig:sdens}} 
\end{figure}

A King model allows an estimate of the spatial number density $\phi(r)$ 
within the cluster.
The surface density in eq. (25) of \citet{king62} was differentiated
numerically. Figure \ref{fig:sdens} shows the resulting spatial densities of
mass groups with the low mass stars subdivided into finer mass groups, for 
stars with $p \ge 0.3$, and assuming a distance modulus of 5.5 \citep{lynga87}. 
The highest mass stars dominate the core in terms of mass density, while 
stars below 1 M$_\odot$ dominate the halo beyond $\sim 5$ pc.  This classical
effect of mass segregation has been well established in the Pleiades and 
other clusters \citep[eg.][]{vanLeeuwen83, hambly95, raboud98b},
and confirms high mass stars in the core are much closer to energy equipartition
than stars in the halo \citep{giersz96, giersz97}. Stars with masses 
less than 1 M$_\odot$ have similar radial extent, implying 
tidal truncation affects stars with mass up to $\sim 1$ M$_\odot$. This result 
indicates that relatively bright stars up to $V \approx 12$ can be used 
to trace escape processes around the cluster using upcoming high precision,
space-based astrometry missions such as FAME \citep{horner99} or SIM \citep{unwin99},
which, according to specifications, should constrain nearly the entire 
phase space for detected stars in nearby clusters such as the Pleiades.

The dynamics that drive mass segregation can deplete low mass objects 
in the core. Simulations \citep[eg.][]{terlevich87} predict spatial 
variations in the mass 
function during core contraction and formation of an extended halo dominated 
by low mass stars. Terlevich's results show a relatively
smaller ratio of 0.3 M$_\odot$ stars to 0.8 M$_\odot$ stars in the core
than in the halo. Thus, migration from the core
can occur preferentially for stars with mass less than the average stellar 
mass in clusters, flattening the slope of the observed mass function at
low mass. 

\begin{figure}
\plotone{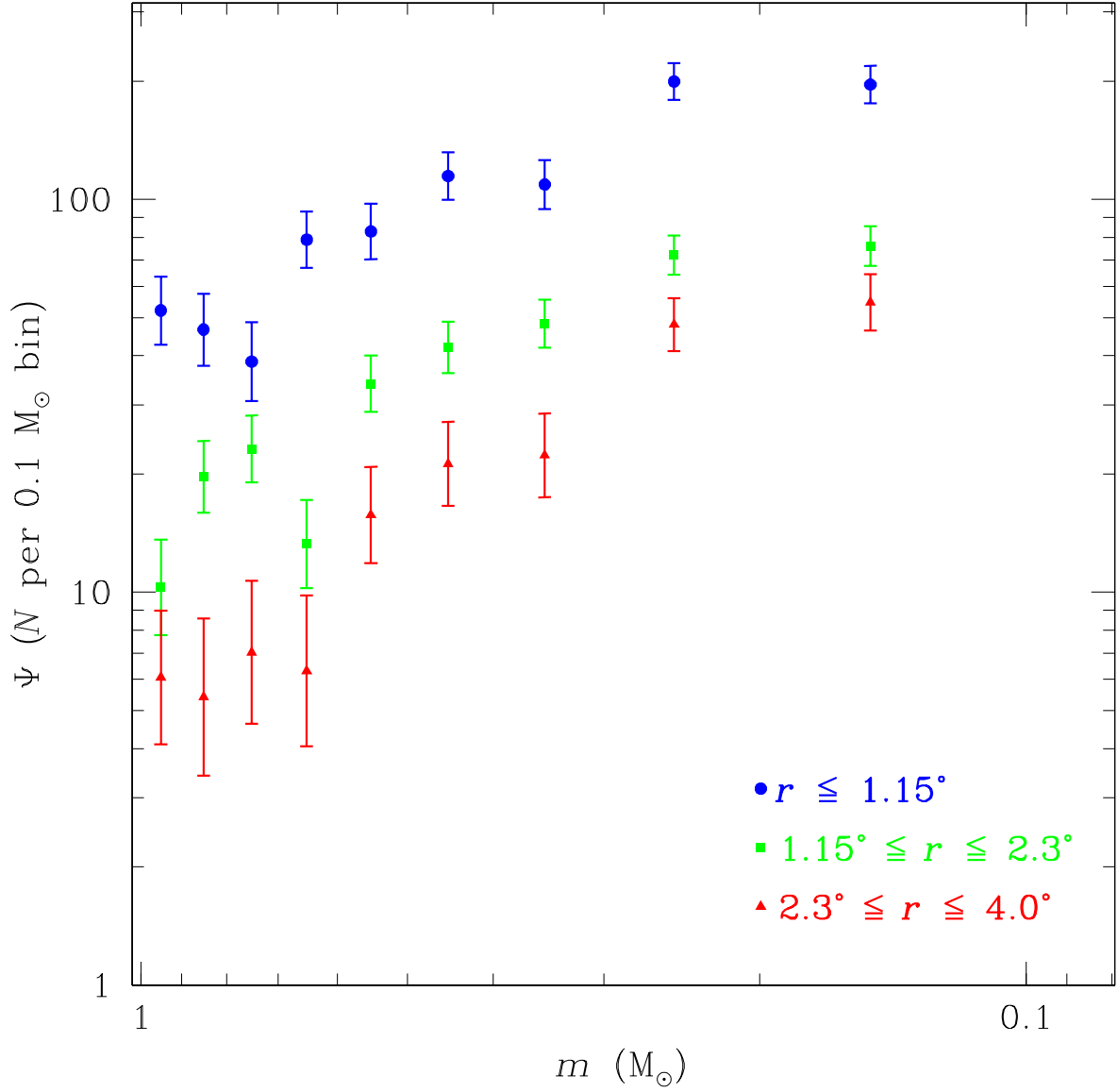}
\caption{Radial dependence of the Pleiades mass function,
corrected for field star contamination. The mass function for
$r \le 1.15^\circ$ has been shifted higher by a factor 2.5
for visual clarity. Error bars represent
Poisson uncertainties. Note 1 core radius is roughly $1.15^\circ$.
\label{fig:mfrad}}
\end{figure}

To look for this effect, we calculated the mass function
$\Psi(m)$, where 
\begin{displaymath}
\Psi(m) = \frac{dN}{dm}
\end{displaymath}
represents the distribution of mass in terms of the number of stars $N$ in the 
mass interval $[m,m+dm]$. Because field star contamination increases with 
radial distance and may pollute the observed mass function, we attempted 
to statistically subtract field contamination. We scaled the number of 
field stars in each mass bin per unit area in the annulus at 
$6.5^\circ - 9.0^\circ$ for the corresponding
area covered by each mass function, and subsequently subtracted the expected
number of field stars from each mass bin. Figure \ref{fig:mfrad} shows the 
dependence of the mass function on radial distance for stars with mass 
1 -- 0.1 M$_\odot$ and $p \ge 0.3$, after field subtraction. The results 
show a more flattened mass function in the core. A chi-square test for binned
data \citep{press92} shows a significant difference in the mass 
functions: there is 11\% probability that the mass function within 1 core 
radius and 1 -- 2 core radii represent the same distribution function,
and 15\% probability the mass function within 1 core 
radius and outside 2 core radii represent the same distribution function.
For comparison, we performed a Kolmogorov-Smirnov test \citep{press92}
between the mass functions {\em before} binning and field subtraction. Results
indicate 21\% probability the mass function within 1 core 
radius and 1 -- 2 core radii represent the same cumulative distribution function,
and less than 1\% probability the mass function within 1 core 
radius and outside 2 core radii represent the same cumulative
distribution function. 

The slightly flatter mass function in the Pleiades core suggests
a marginal detection of preferential depletion of low mass objects
at the center of the cluster. Moreover, projection of foreground halo
stars onto the observed core means the mass function in the central part
of the cluster may be more depleted of low mass objects than observed in
projection. Tidal truncation below $\sim 1$ M$_\odot$
then implies slightly higher relative escape rates for stars below 
average mass, roughly 0.4 -- 0.5 M$_\odot$. Preferential escape may affect
the slope of the overall observed mass function, discussed next.

\subsubsection{Overall mass function \label{sec:mf}}

The abundance of low mass stars in stellar systems makes them important
players in dynamical processes. The character of the mass function (MF)
in rich, young clusters such as the Pleiades has important ramifications for 
the initial mass function (IMF) in star formation, since most stars are 
believed to form in clusters \citep{lada93}. The MF is also a critical 
ingredient in 
$N$ body simulations with implications for the internal dynamics of energy
equipartition \citep[eg.][]{inagaki85, giersz96} and expected
lifetime of clusters \citep{terlevich87, delafuente95}. 

\begin{figure}
\plotone{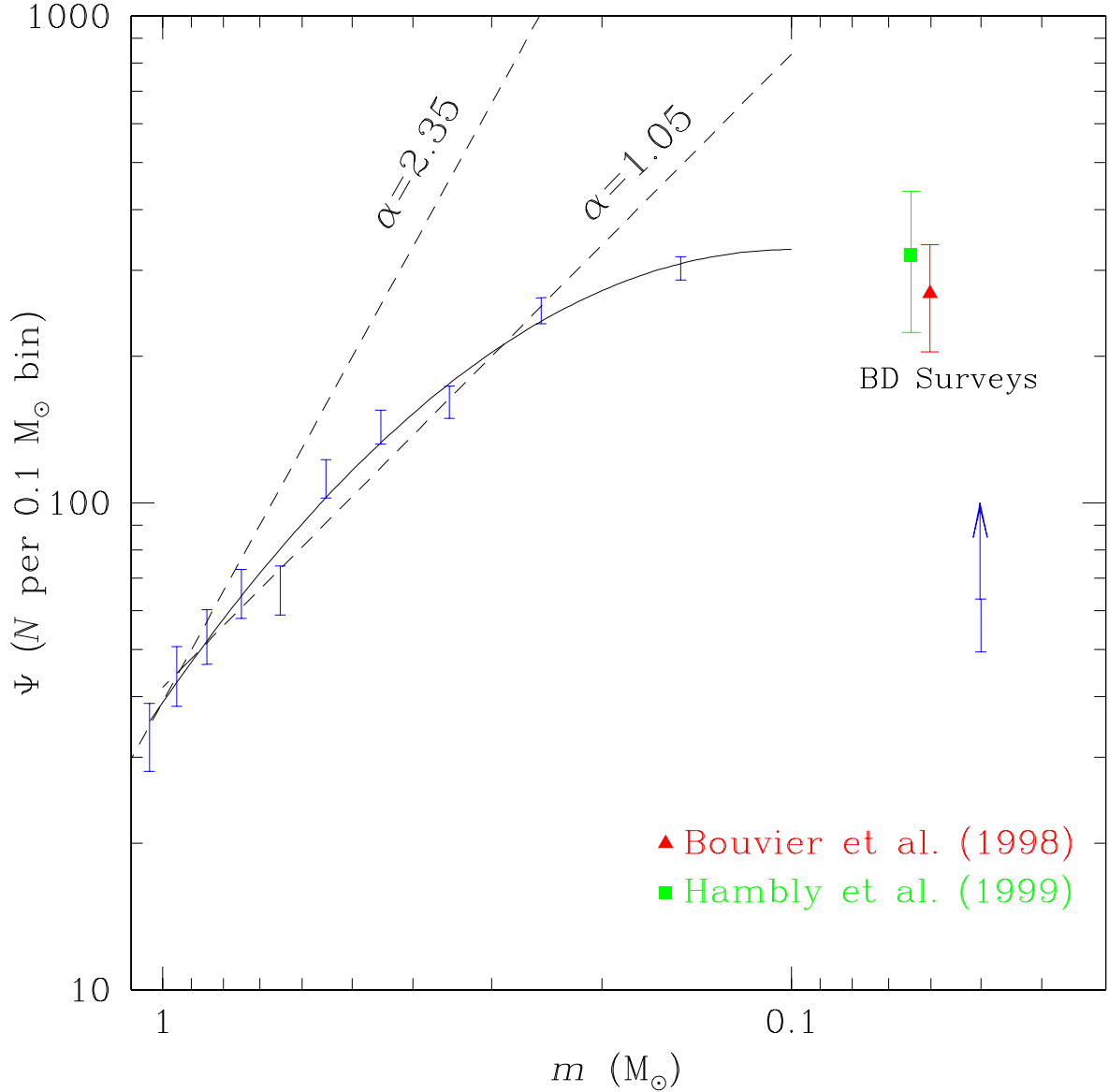}
\caption{Pleiades mass function below 1 M$_\odot$. The solid line indicates
the best fit polynomial to 0.1 M$_\odot$ where the sample becomes 
incomplete. Also shown are results from the brown dwarf surveys of 
\citet{bouvier98} (solid triangle) and \citet{hambly99} (solid box).
For comparison, the dashed lines plot power law mass functions
for the Salpeter case ($\alpha=2.35$) and best fit ($\alpha = 1.05$)
to the local field from 1 -- 0.1 M$_\odot$ \citep{reid97}, normalized
to $\sim 1$ M$_\odot$.
\label{fig:mf}}
\end{figure}

Figure \ref{fig:mf} shows the distribution 
of Pleiades masses below 1  M$_\odot$ for stars with $p \ge 0.3$ and
radial distance less than $5.5^\circ$. Error bars portray Poisson
statistics.  The completeness of the sample begins
to decrease below $\sim$ 0.15 M$_\odot$. We agree with \citet{hambly99} that the 
Pleiades MF is best fit in $\log dN/dm \: vs. \: \log m$ space
with a log normal (quadratic) function rather than a single exponent power law. 
The best polynomial fit in the 1 -- 0.1 M$_\odot$ region gives
\begin{displaymath}
\log (N \: \rm{per \: 0.1 \: M}_\odot \: \rm{bin}) = -0.93 \pm 0.03(\log {\it m})^2 - 1.86 \pm 0.02 \log {\it m} + 1.59 \pm 0.02
\end{displaymath}
as delineated in Figure \ref{fig:mf}. 
The overall field star contamination in this sample is expected to be roughly 
10 -- 15\% (or less for stars above 0.5 M$_\odot$). This amount of field 
star contamination should have little effect on the coefficients of the fit. 
Thus the Pleiades MF appears to be slowly rising
below 0.5 M$_\odot$ and relatively flat at 0.1 M$_\odot$. This MF could be 
considered a lower limit at the faint end due to incompleteness and 
unresolved binarity. 

Although less steep in the 0.3 -- 0.1 M$_\odot$ region than those inferred by
brown dwarf surveys \citep{bouvier98, hambly99}, this Pleiades
MF appears consistent with previous results, as shown in Figure \ref{fig:mf},
and suggests a flat or decreasing MF at the hydrogen burning limit. 
Assuming that this MF continues below 0.04 M$_\odot$, the 303 objects in the 
lowest 0.1 M$_\odot$ mass bin imply a {\em crude} lower limit of 
$\sim 200$ free-floating objects below the hydrogen burning mass limit
at 0.075 M$_\odot$. Unless the MF rises sharply below 0.06 M$_\odot$,
the total brown dwarf mass contribution is likely to be less than 5\% of 
the total Pleiades mass. 

For comparison, Figure \ref{fig:mf} also shows power law
functions, $\Psi(m) \propto m^{-\alpha}$, for a Salpeter 
MF ($\alpha = 2.35$) \citep{salpeter55} and for the local field MF 
($\alpha = 1.05$) \citep{reid97}, normalized to $\sim 1$ M$_\odot$.
While our MF alone is not necessarily
inconsistent with $\alpha \approx 1$ at 0.15 M$_\odot$ due to incompleteness
and binarity (discussed below), our results combined with those of 
\citet{bouvier98} and \citet{hambly99} seem to show a persistent flattening below 
0.2 M$_\odot$, in contrast to the local field MF, which continues to rise to 
0.1 M$_\odot$. 

A key question is whether the Pleiades MF represents initial conditions
or dynamical evaporation. Substantial evaporation will flatten the present
day MF from the IMF, since low mass objects escape preferentially. 
However, the Pleiades may have evaporated too few stars
to alter the slope of the observed IMF at low mass.
For example, \citet{delafuente00} performed simulations which showed 
little fractional depletion of brown dwarfs, compared with stars, 
until cluster half-life. Without better membership information at large
radial distances, and a reliable estimate of the number of escaped cluster 
members, we cannot quantify the precise degree of evolution the MF exhibits. 

\subsubsection{Unresolved binarity}

Many Pleiades proper motion candidates are unresolved binaries that will 
systematically affect measurements of the mass function and
total mass in stars. In order to derive a correction for unresolved
binaries, we performed a simple simulation of the observed masses in a cluster 
with known primary mass function and frequency and mass distribution of binary 
companions. The simulation used standard Monte Carlo techniques, 
as well as the mass-luminosity relations mentioned in \S\ref{sec:merge}.

The primary mass distribution consisted of a generic log-normal MF. 
A binary frequency function $f(m)$, and a 
distribution of mass ratios $q(m) = m_2/m_1$ where $m_1$ is the primary
mass and $m_2$ the secondary mass, determined secondary masses.
The hypothetical cluster contained stars in high, medium, and low mass groups
of $m/\rm{M}_\odot \ge 2$; $1 \le m/\rm{M}_\odot < 2$; 
and $m/\rm{M}_\odot < 1$, respectively. 

Realistic values for $f(m)$ are 0.5 for the high and medium
mass group and 0.3 for the low mass group \citep{abt76, mermilliod92}.
Input parameters for the mass function and $q$ distribution
were adopted from \citet{duquennoy91} and \citet{kroupa90}. 
The primary masses were distributed in N (per unit log$m$) -- log$m$ space 
with gaussian peak at $\log m \sim -0.5$ and width of $\sim 0.4$. 
Since they are poorly determined, we varied the 
parameters of the MF distribution slightly during the simulations, typically by 
a couple of tenths in central value and width. The $q$ distribution consisted
of a gaussian with peak 0.23 and width 0.42. For stars with mass lower than 1 
M$_\odot$, the $q$ distribution was flattened, as suggested by data from
\citet{mermilliod92}. 

\begin{figure}
\plotone{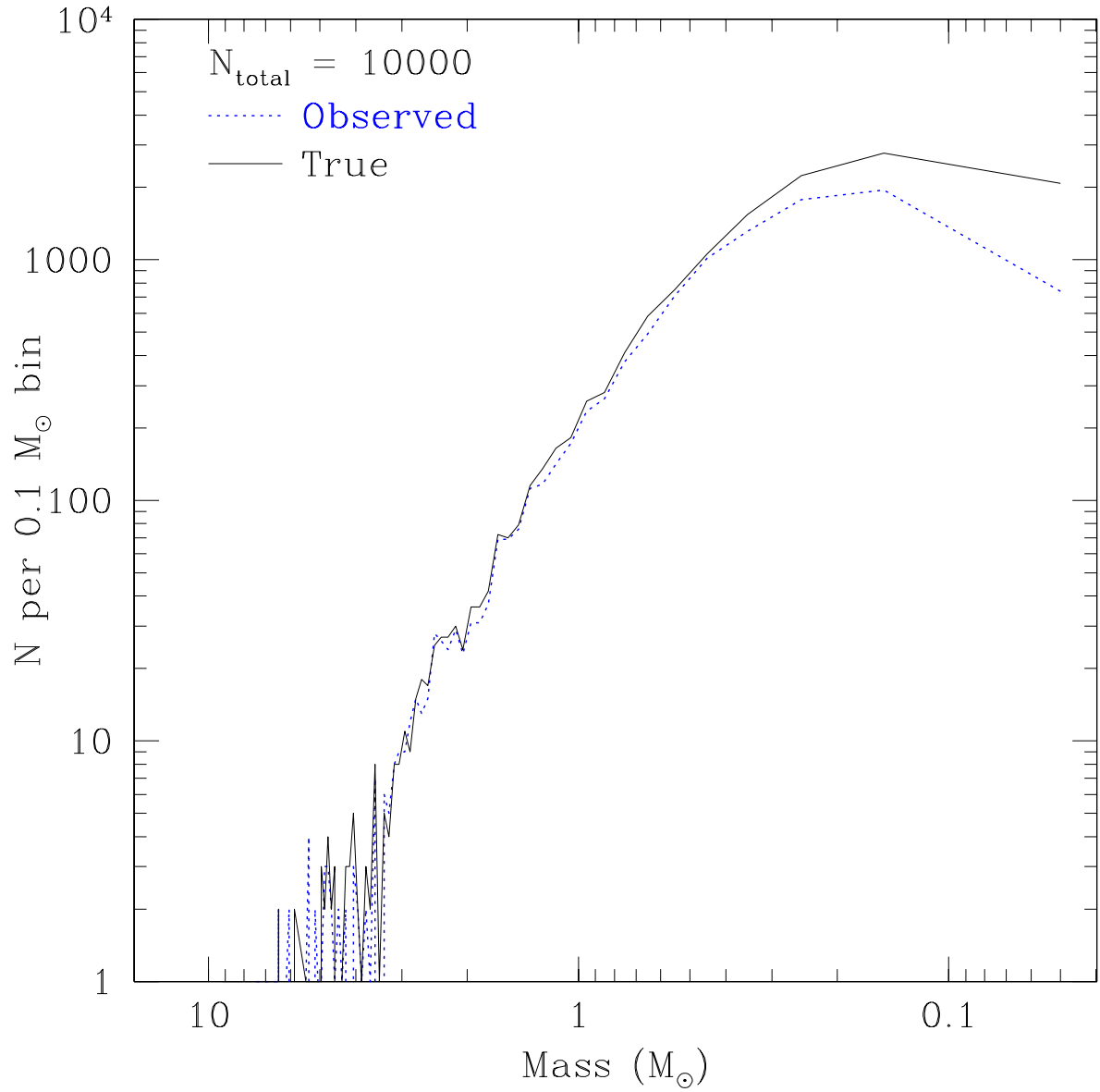}
\caption{Simulated mass functions for a cluster of 10000 stars. In the observed 
case, all binary systems are unresolved.
\label{fig:mfsim}}
\end{figure}

Figure \ref{fig:mfsim} shows a typical observed mass function for a given true mass 
function in $\log dN/dM - \log m$ space for a cluster containing 10000 stars.
The observed mass function underestimates the true mass function 
below 0.5 M$_\odot$ by a factor of 1 -- 2 down to 0.1 M$_\odot$. 
This simulation assumed that all Pleiades binaries were unresolved,
since 2MASS spatially resolves only the widest binaries ($> 10^5$ AU) at the 
distance of the Pleiades. The simulations suggest that unresolved binaries at 
low mass may affect the slope of a mass function $\alpha$ by up to a few 
tenths below 0.5 M$_\odot$. Repeated simulations determine an overall mass correction
factor $\delta$ where $M_{true} = (1+\delta) M_{obs}$.
Computing the effective apparent magnitude for
the binary stars, and integrating the true and observed mass functions to about
0.08 M$_\odot$, gave $\delta \sim$ 0.15. The simulations were most sensitive to $f(m)$.
If $f(m)=1$ for all masses, $\delta$ increases to $\sim$ 0.3. 
Sliding the peak of the mass function 0.2 $\log m$ units resulted in a 
change in $\delta$ by about 0.02, where a higher peak causes a
larger $\delta$. A change the width
of the mass function by 0.2 $\log m$ units also resulted in a change in
$\delta$ by about 0.02, where a wider mass function causes a larger
$\delta$. Note that $\delta$ was similar for repeated 
simulations of a cluster of 1000 stars. 

This simulation is obviously crude, and eg. SIRTF observations would better
determine Pleiades binarity at low mass. 

\subsection{Dynamical mass}

A simple sum of the masses of individual members yields the total 
Pleiades mass. Candidates in the merged sample with $p \ge 0.3$ within a 
5.5$^\circ$ radial distance from the nominal center are the most probable 
members. Integration of the individual masses to 0.1 M$_\odot$ gives a total 
mass of 690 M$_\odot$ and a binary-corrected mass of $\sim$ 800 M$_\odot$. 
This simple sum contains systematic and random errors from 
field contamination, unresolved binaries, Poisson uncertainties, and completeness. 
Objects below 0.1 M$_\odot$ are not likely to add more than a 
few percent to this mass estimate. 

\begin{figure}
\plotone{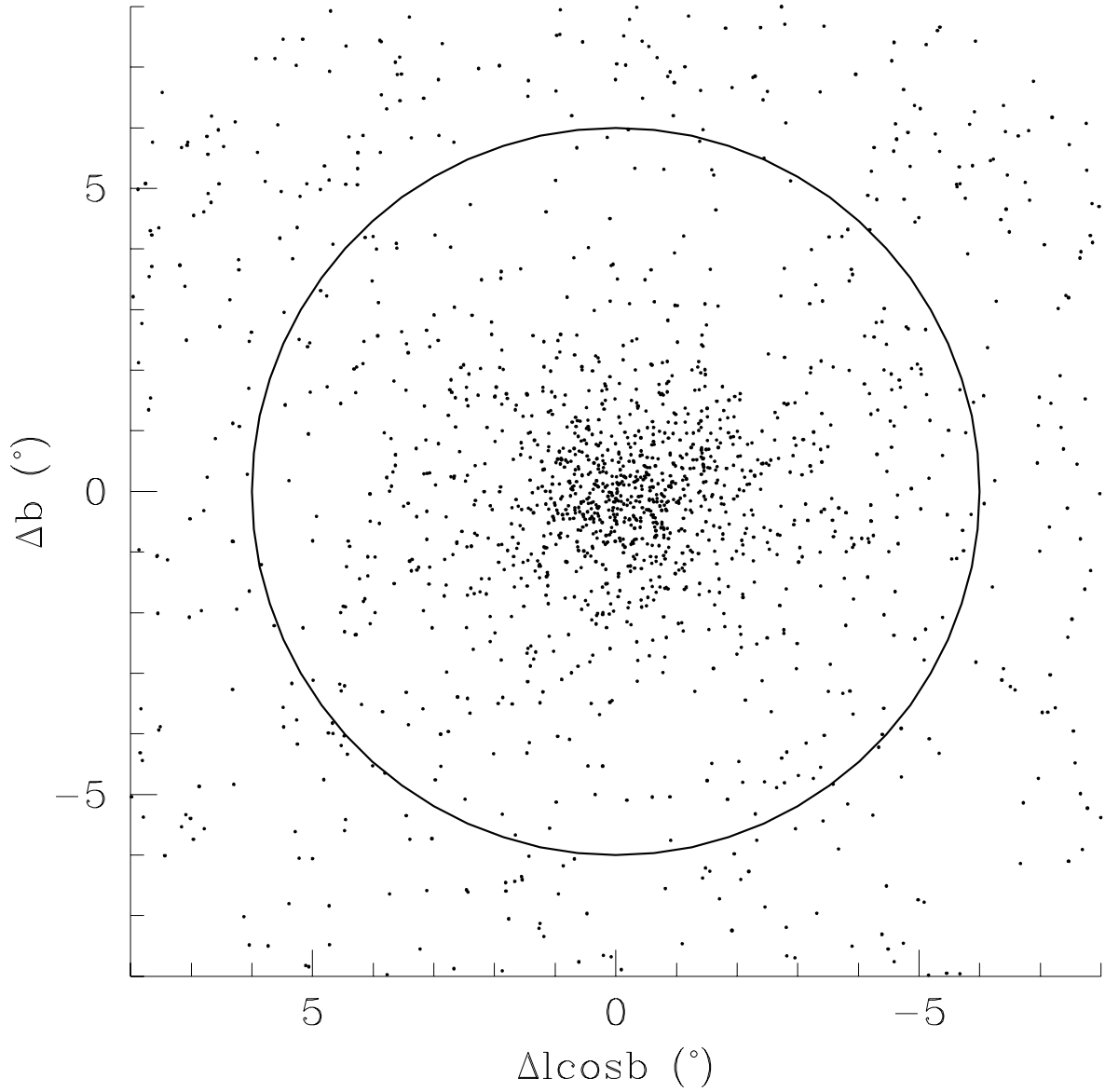}
\caption{Spatial distribution in Galactic coordinates for 1 -- 0.1 M$_\odot$
Pleiades candidates. The circle traces the projection of the tidal radius 
derived from a total mass estimate of $\sim$ 800 M$_\odot$. Nearly all 
objects outside the tidal radius in this plot belong to the field.
\label{fig:spatial}}
\end{figure}

For a local Galactic cluster in circular orbit, the approximate limiting 
tidal radius $r_{lim}$ for a member star along the direction towards or away 
from the Galactic center depends on the cluster mass $M_c$ and the Galactic
rotation curve \citep{king62}:
\begin{displaymath}
r_{lim} = \left [ \frac{GM_c}{4 A (A-B)} \right ]^\frac{1}{3}
\end{displaymath}
where $A$ and $B$ are the Oort constants.
The values $A=14.4$ km s$^{-1}$ kpc$^{-1}$ and $B=-12.0$ km s$^{-1}$ kpc$^{-1}$
\citep{kerr86} correspond to a tidal radius of 13.1 pc, which confirms earlier
results found by \citet{pinfield98} using several different Pleiades catalogs
and an expression for the tidal radius of a cluster in radial orbit
\citep{vonhoerner57}.
Figure \ref{fig:spatial} displays a projection of this radius. The tidal radius 
overlaps with the confidence interval for the ``edge'' of the cluster determined
from the King profile fit (Figure \ref{fig:fsall}). Therefore, the limited
size of the Pleiades indicates the absence of a massive population of 
brown dwarfs or internal dark matter. Thus we conclude that the
stellar mass composes nearly the entire dynamical mass. 
Further comparison of $r_{lim}$ given above with the 
King tidal radius $r_t$ is dangerous as they represent different quantities in 
this case, since the direction of $r_{lim}$ lies along the line of sight.

\subsection{Projected halo structure}

$N$ body models predict an elongation of open clusters in roughly the 
$X$-direction and to a slightly lesser extent in the $Y$-direction, due to 
heating in the halo by the Galactic tidal field \citep{terlevich87, portegies00}. 
The tidal field in these models, generated from a 
circular cluster orbit in the Galactic plane, produces a nonspherical
tidal surface resulting in a flattened structure with axial ratios 2:1.4:1 in the 
$X$:$Y$:$Z$ directions, respectively \citep{wielen75, aarseth73}. 
Terlevich predicts a fairly constant number of stars trapped in the halo, 
between 1 and 2 tidal radii, for some period of 
time before escaping permanently. Portegies Zwart maps escaping stars 
exiting primarily through the first and second Lagrangian points.

\begin{figure}
\plotone{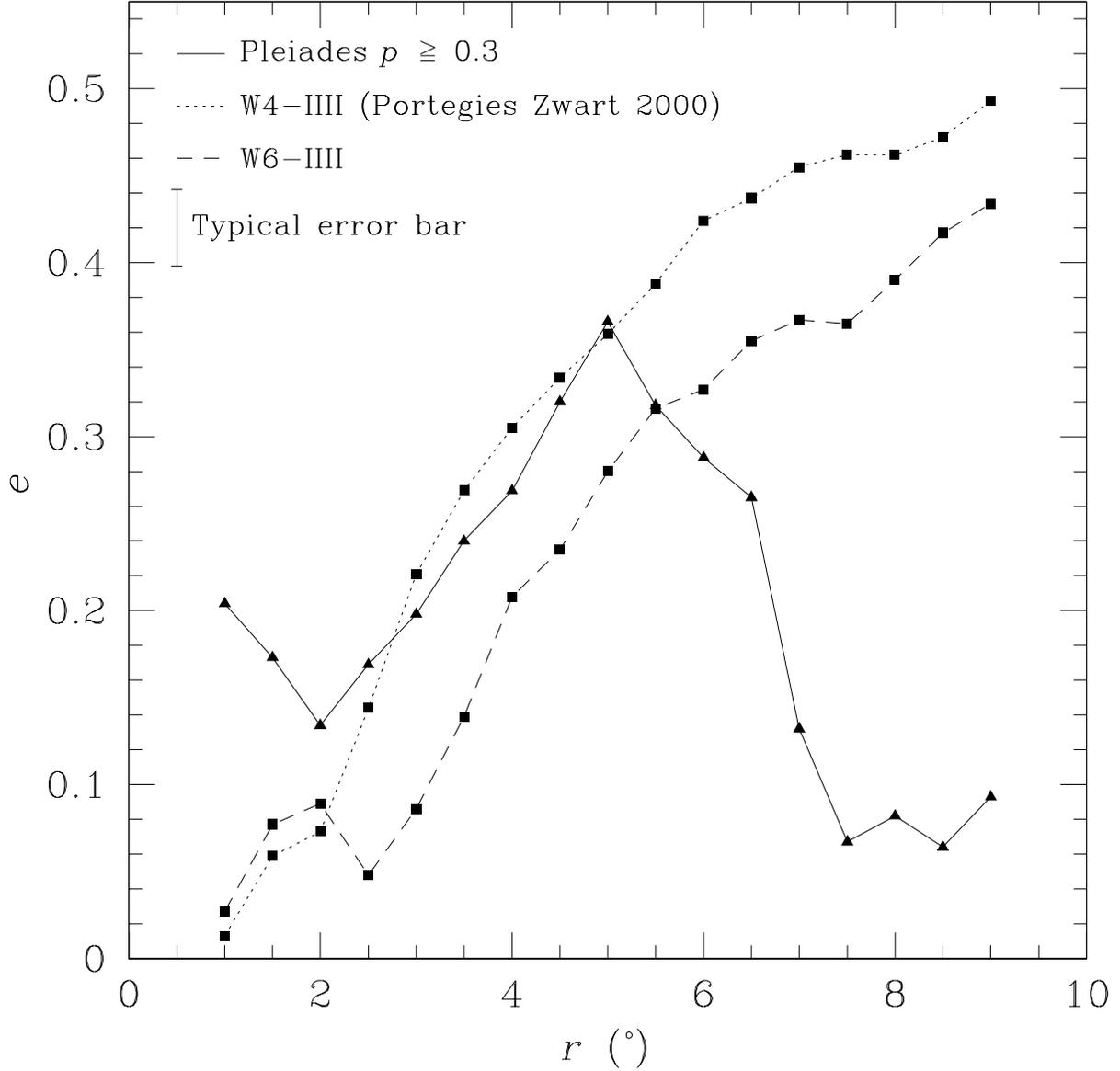}
\caption{Ellipticity $e$ of the Pleiades, plotted cumulatively {\it vs.} radial distance 
$r$, for stars with mass 1 -- 0.1 M$_\odot$ and $p \ge 0.3$. Dotted and broken and lines 
indicate analogous measurements for models W4-IIII and W6-IIII from \citet{portegies00},
simulated at 6 and 12 kpc from the Galactic center, at ages of 100 Myr. The simulated
clusters have been scaled and projected onto the sky as if they were at the
distance of the Pleiades. The error bars were estimated using Poisson uncertainties; only
the typical error bar is shown for visual clarity. 
\label{fig:e}}
\end{figure}

At $l=166^\circ$ and $b=-23$, the projection of the Pleiades onto the sky is 
most sensitive to structure in the $YZ$ plane. Following \citet{raboud98a},
we used moment analysis to determine the ellipticity $e = (1-b/a)$ of the 
Pleiades. This method allows direct, quantitative comparison of the Pleiades 
to a simulated cluster.
Figure \ref{fig:e} shows measurements of $e$ within a given radius $r$
for Pleiades stars with mass 1 -- 0.1 M$_\odot$, with propagation of
Poisson uncertainties determining the error bars. Field contamination
dilutes the measured ellipticity, especially near and beyond the tidal
radius at $\sim 6^\circ$. For comparison, Figure \ref{fig:e} also shows
the ellipticity of Portegies Zwart's \citeyearpar{portegies00}
models W4-IIII and W6-IIII at an age of 100 Myr. The W4 and W6 simulated 
clusters orbit at 6 kpc and 12 kpc,
respectively, from the Galactic center. To compare their ellipticities
with the Pleiades, we projected the positions of stars with mass 1 -- 0.1
M$_\odot$ to the location and size of the Pleiades on the sky, assuming
binaries are unresolved. The models show similarity to the flattened structure 
throughout most of the Pleiades halo. The moment analysis confirms elongation in 
both the Pleiades and the models lies nearly parallel the Galactic plane;
the major axes of the halos of the Pleiades and models lie consistently
within $10^\circ$ of the longitude of the cluster. However, the orientation
of the ellipticity in the core of the Pleiades deviates more randomly from the halo 
by as much as $30^\circ$, suggesting the higher ellipticity in the
Pleiades core seen in Figure \ref{fig:e} is probably due to coincidental projection 
effects rather than a consistent elliptical distribution, and thus is not
physically significant. 
Clearly, better membership information is required for the outermost Pleiades stars. 
However, the models appear to adequately describe a young, but dynamically evolved, 
open cluster such as the Pleiades. 

The Pleiades cluster has interacted with local environmental features not specifically 
included
in the models. For example, the Pleiades probably migrated from the Sagittarius arm 
\citep{yuan77}. However, this would have occurred more than one relaxation
time interval ($\sim 50$ Myr) in the past, so the cluster, through evolution, has 
likely ``forgotten'' any effects from its passage out of the arm. 
Recent observations of interstellar gas in the vicinity of the
Pleiades suggests that the cluster is coincidentally colliding with 
two small molecular clouds \citep[eg.][]{white93,white97} moving, relative to
the Pleiades, at $\sim$ 15 -- 20 km/s. However, the mass of the gas is only 
20 -- 50 M$_\odot$ \citep{bally86,breger87}, which is much 
smaller than the mass of the cluster stars. Thus, comparison of the Pleiades with models
is legitimate since local features have had negligible effects on its present
structure.

The Poisson statistics of the background stars in Figure \ref{fig:spatial}
place an upper limit of $\sim 1.7$ stars deg$^{-2}$ that may have escaped the Pleiades 
in the projected $YZ$ plane. Without individual distance measurements, projection of 
the $X$ direction along the line of sight prevents our probing the proposed escape 
mechanism near the Lagrangian points. 

Since the Galactic tidal field is the primary catalyst for dispersal,
a future measurement of the escape mechanism combined with structure and kinematic
data in open clusters of various dynamical ages will better test our understanding of 
the nature of the tidal field. For example,
substantial flattening has also been found in the Hyades \citep{oort79, perryman98} 
and Alpha Persei \citep{prosser92}. However, other
clusters, such as Praesepe \citep{raboud98b} and M67 \citep{mathieu83}, show 
less flattening. Better membership surveys are required in most clusters 
to study any dependence of structure on age or environmental effects. 
A comparison of the static tidal model results to those 
of more realistic simulations that include time dependent tidal fields from radial 
oscillation and vertical disk passage \citep[eg.][]{murali97, combes99} would be 
a useful exercise to closely examine when the static approximation to the tidal
field breaks down in clusters older than the Pleiades.

\section{SUMMARY}

This paper described a  wide field proper motion search using 2MASS and PMM astrometry,
which has detected $\sim$ 1200 high probability, low mass Pleiades candidates,
down to a mass $\sim$ 0.1 M$_\odot$, from nearly 1 million stars in a 300 deg$^2$ 
field around the nominal center of the cluster. Several hundred are newly detected 
candidates. Spectroscopic results for 528 Pleiades candidates show a slight degree of 
contamination for high probability candidates to a radial distance of $\sim$ 5$^\circ$. 
Beyond this distance, nearly all objects detected as Pleiades candidates are expected 
to be field stars.

The high mass cluster members are centrally concentrated with respect to 
low mass members. Tidal truncation is relevant to all objects with mass less
than $\sim 1$ M$_\odot$. The mass function in the range 1 -- 0.1 M$_\odot$
is marginally flatter within the core than
outside, indicating evidence for preferential escape of stars with
less than average mass. These results agree with 
theoretical and numerical expectations.

The Pleiades mass function rises slowly below 1 M$_\odot$ and appears to become 
flat at 0.1 M$_\odot$, suggesting a lower limit of $\sim$ 200 brown dwarfs
in the cluster. This result is consistent with previous surveys that conclude
brown dwarfs contribute little to the dynamical mass. Stars with mass 
1 -- 0.1 M$_\odot$ dominate the total mass. Better membership information, near
and beyond the tidal radius, is 
needed to determine whether the Pleiades mass function represents the initial mass 
function, or reflects preferential evaporation of low mass objects.
The total Pleiades stellar mass is $\sim$ 800 M$_\odot$, including
a 15\% correction for unresolved binaries. The corresponding tidal 
radius to this mass estimate is approximately 13.1 pc. The
apparent size of the Pleiades rules out a significant addition to this 
mass estimate from brown dwarfs or dark matter. 

The cluster is highly flattened in the $YZ$ plane, in good agreement
with recent numerical models. This result confirms
that the Galactic tidal field plays the dominant role in open cluster dispersal.
Detection of the escape mechanism and dispersal rate will likely rely on
future high precision, space-based astrometry.

\acknowledgments
We thank R. Cutri for assistance with 2MASS point source files for the Pleiades 
field. J. A. gratefully acknowledges S. Portegies Zwart for sharing and discussing
simulation results. M. Weinberg and R. White provided helpful criticism for parts of 
this manuscript. NOAO funds for graduate student thesis observations provided 
travel support for J. A. This publication makes use of data products from the 
Two Micron All Sky Survey, which is a joint project of the University of Massachusetts 
and the Infrared Processing and Analysis Center, funded by the National Aeronautics and 
Space Administration and the National Science Foundation. The
National Geographic Society-Palomar Observatory Sky Atlas (POSS I) was made by the 
California Institute of Technology with grants from the National Geographic Society. 
The Second Palomar Observatory Sky Survey (POSS II) was made by the California 
Institute of Technology with funds from the National Science Foundation, the National
Geographic Society, the Sloan Foundation, the Samuel Oschin Foundation, and the 
Eastman Kodak Corporation. This research made use of the Digitized Sky Survey. 
The Digitized Sky Survey was produced at the Space
Telescope Science Institute under US government grant NAGW-2166. The images of these 
surveys are based on photographic data obtained using the Oschin Schmidt Telescope on 
Palomar Mountain and the UK Schmidt Telescope. The plates were processed into the 
present compressed digital form with the permission of these institutions.
This research has made use of NASA's Astrophysics Data System Abstract
Service.

\newpage

\newpage

\newpage

\newpage

\newpage

\newpage

\newpage

\newpage

\newpage

\newpage

\newpage

\newpage

\newpage

\newpage

\newpage
\begin{deluxetable}{cccrrc}
\tabletypesize{\small}
\tablecaption{\label{tab:wiynfields} Coordinates for WIYN Hydra spectrograph fields with the number of Pleiades candidates, field control sources, and exposure time.}
\tablehead{\colhead{Date (2000 UT)} & \colhead {RA (J2000)} & \colhead {DEC} & \colhead {Candidates} & \colhead {Control} & \colhead {Exp. time (min.)}}
\startdata
Nov 26 02:48 & 03 44 50 & +24 32 47 & 43 &  0 & 90 \\ 
Nov 26 05:23 & 03 44 54 & +23 36 20 & 33 &  4 & 90 \\ 
Nov 26 08:14 & 03 47 00 & +24 06 00 & 49 &  3 & 90 \\ 
Nov 26 10:34 & 03 49 01 & +23 39 56 & 23 &  3 & 90 \\ 
Nov 27 02:42 & 03 42 53 & +24 03 53 & 48 &  4 & 90 \\ 
Nov 27 04:47 & 03 47 07 & +23 11 31 & 37 &  6 & 90 \\ 
Nov 27 07:44 & 03 51 09 & +24 09 15 & 20 &  1 & 90 \\ 
Nov 27 09:49 & 03 48 50 & +24 37 33 & 37 &  2 & 90 \\ 
Nov 27 11:56 & 04 00 04 & +23 49 00 & 17 &  6 & 30 \\ 
Nov 28 02:38 & 03 50 49 & +25 05 01 & 18 &  2 & 90 \\ 
Nov 28 04:40 & 03 51 08 & +23 11 37 & 13 &  6 & 90 \\ 
Nov 28 07:13 & 03 43 06 & +23 07 36 & 35 &  6 & 60 \\ 
Nov 28 08:44 & 03 44 26 & +25 29 21 & 31 &  6 & 60 \\ 
Nov 28 10:54 & 03 58 24 & +21 42 00 & 12 &  5 & 60 \\ 
Nov 29 02:39 & 03 31 40 & +26 12 00 & 10 &  5 & 45 \\ 
Nov 29 03:56 & 03 34 48 & +25 24 00 & 12 & 12 & 45 \\ 
Nov 29 05:47 & 03 40 24 & +25 27 33 & 31 &  9 & 45 \\ 
Nov 29 07:00 & 03 49 24 & +21 47 05 & 13 &  9 & 45 \\ 
Nov 29 08:54 & 03 54 00 & +19 42 00 &  7 &  9 & 30 \\ 
Nov 29 09:54 & 03 38 47 & +23 59 57 & 31 & 11 & 45 \\ 
Nov 29 11:45 & 03 56 24 & +26 42 00 &  8 &  4 & 45 \\ 
\enddata
\end{deluxetable} 

\newpage
\begin{deluxetable}{crrrrrrr}
\tabletypesize{\small}
\tablecaption{\label{tab:Hstats} Number of Pleiades candidates distributed in $W_{H\alpha}$ (\AA) bins according to membership probability $p$, $K_s$ magnitude, and radial distance $r$ from the cluster center.}
\tablehead{\colhead { } & \colhead {0-1 \AA} & \colhead {1-3 \AA} & \colhead {3-5 \AA} & \colhead {5-7 \AA} & \colhead {7-9 \AA} & \colhead {9-11\AA} & \colhead {$> 11$ \AA}}
\startdata
\cutinhead{$p$}
$0.001 \le p \le 0.01$  &  22  &   0  &   3  &   5  &   1  &   0  &   0 \\
$0.01 \le p \le 0.1$  &  19  &   3  &  10  &  18  &  10  &   3  &   3 \\
$0.1 \le p \le 0.3$  &  11  &   6  &  12  &  17  &  17  &   6  &   4 \\
$0.3 \le p \le 1.0$  &  27  &   8  &  56  & 131  &  98  &  29  &   9 \\
\cutinhead{$K_s$}
$11 \le K_s \le 12$ &   3 &   2 &  14 &  27 &  15 &   1 &   0 \\
$12 \le K_s \le 13$ &  33 &   2 &  17 &  85 &  69 &  23 &   7 \\
$13 \le K_s \le 14$ &  29 &   8 &  46 &  55 &  38 &  14 &   7 \\
$14 \le K_s \le 15$ &  14 &   5 &   4 &   4 &   4 &   0 &   2 \\
\cutinhead{$r$}
$0^\circ \le r \le 1^\circ$ &  14 &   5 &  33 &  84 &  62 &  20 &   8 \\
$1^\circ \le r \le 2^\circ$ &  35 &   6 &  31 &  61 &  48 &  11 &   6 \\
$2^\circ \le r \le 3^\circ$ &  11 &   2 &  11 &  14 &  12 &   5 &   0 \\
$3^\circ \le r \le 4^\circ$ &  15 &   3 &   4 &  10 &   3 &   2 &   2 \\
$4^\circ \le r \le 5^\circ$ &   3 &   1 &   2 &   2 &   1 &   0 &   0 \\
$5^\circ \le r \le 6^\circ$ &   1 &   0 &   0 &   0 &   0 &   0 &   0 \\
\enddata
\end{deluxetable} 

\end{document}